# Scanning Tactile Sensor with Spiral Coil Structure Amplifying Detection Performance of Micro-concave

Takeru Kurokawa[1], Toshinobu Takei[3], Haruo Noma[2] and Mitsuhito Ando[2]

[1] Graduate School of Information Science and Engineering, Ritsumeikan University, Osaka, Japan

[2] College of Information Science and Engineering, Ritsumeikan University, Osaka, Japan

[3] Faculty of Science and Technology, Seikei University, Tokyo, Japan

(Email: tkurokawa@mxdlab.net)

**Abstract ---** Surface inspection is a delicate process aimed at detecting fine defects, irregularities, and foreign substances at the tens of micrometers level, subsequently excluding products that do not meet quality standards as defective. Currently, this inspection relies on the tactile senses of skilled technicians, leading to variability in the detection accuracy based on the level of proficiency and experience. Consequently, a standardized method for surface inspection has yet to be established. In response to this issue, we developed a device capable of amplifying tactile information, allowing for the detection of minute distortions without the need for highly skilled technicians. The experimental results for various small distortions suggest the potential for the quantitative evaluation of these distortions. In the future, the application of this device could contribute to the automation of surface inspection.

## 1 INTRODUCTION

In the automotive production lines, surface inspection is an important operation for maintaining product quality in automotive production lines. The inspection is used to detect micro bumps of several tens of micrometers, which are defects that exist on the surface of products. Currently, surface inspection is performed by skilled workers who trace on the surface by hand. Because it relies on the human sense of touch, the detection rate of micro bump varies depending on individual differences and physical conditions. Therefore, automation of surface distortion inspection is required for more efficient production. For automation, sensing of micro bumps is necessary. We have proposed a sensing method using the tactile amplification structure of micro bumps. In our previous research, we developed a scanning tactile sensor using a tactile amplification structure called Touch Lens [1-2] as the sensor structure [3]. In this sensor, the tactile amplification structure follows the surface, amplifying the information of the micro-surface shape and transmitting it to the sensor element. However, it is difficult to make the sensor follow a concave surface and to detect minute concavities. The purpose of this research was to sense micro bumps at the level of several micrometers without using sensory evaluation.

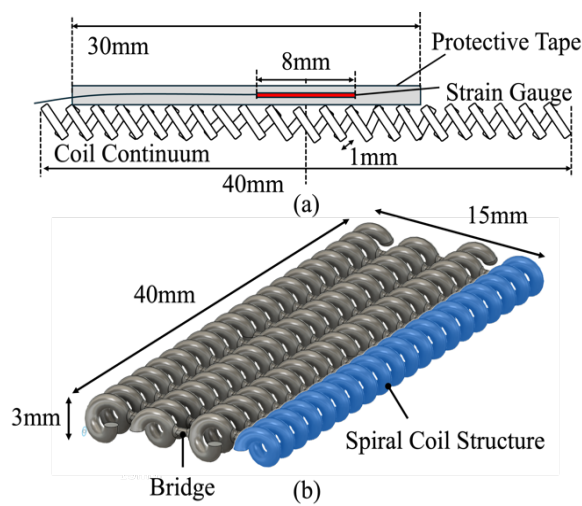

Fig.1 Comparison of the proposed device. (a) The proposed device, (b) the structural components of the device

Surface inspection with optical sensing has been proposed in research on surface inspection [4-5]. However, errors on the order of several micrometers can easily occur because of vibration and contamination of the bottom plate on which the object is fixed, resulting in misrecognition. Therefore, when using optical sensing, it is necessary to minimize vibrations and clean the surface before inspection.

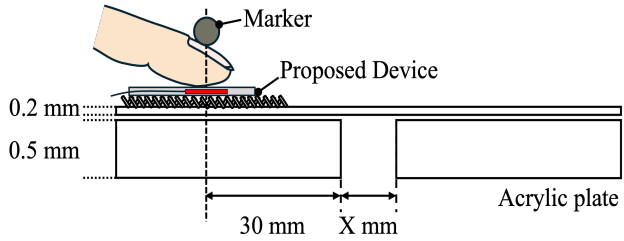

Fig.3  Geometry of tracing surface

However, it is difficult to apply this method to a production line in which quick inspection is required. In addition, sensors have been developed to sense minute surface features by pressing them against a surface [6-7]. Although these sensors can measure surface profiles with high accuracy, the time required to measure a wide area, such as a car body, may be longer than that of conventional methods because of the need to press the sensor against the surface to be measured. Our proposed sensing system [3] solves these problems by employing a surface-tracing method to detect micro bumps.

With respect to concavity, we developed a spiral coil structure as a tactile amplification structure [8]. This structure is a spiral coil wound on a single- or multi-layered plane, and its elasticity in the longitudinal direction allows it to be easily bent and deformed. Owing to its flexibility, it can be deformed along the smallest indentation, thereby amplifying tactile sensation. In this study, we propose a new tactile device based on a spiral coil structure. In a previous study [8], the detection of indentations with a depth of 5 μm was successfully demonstrated. In this study, we investigate the impact of varying strain parameters on detection performance.

## 2  A NOVEL DEVICE WITH TACTILE AMPLIFICATION EFFECTS

The developed device is shown in Fig. 1(a). The device consists of a sensor part and a structural part. When the structural part passes over an indentation, it deforms significantly, and the sensor part detects this deformation to identify minute distortions. The sensor part employs a strain gauge covered with stretchable protective tape.

The structural part features a "coil continuum" that applies the tactile amplification effect of the spiral coil structure, as shown in Fig. 1(b). This structure consisted of spiral coils connected continuously in the transverse direction. To effectively utilize the characteristics of the spiral coil structure, each adjacent coil was connected to a bridge, allowing for flexible movement in the longitudinal direction without interference. The spiral coils were fabricated with a diameter of 1 mm, whereas bridge sections were created with a diameter of 0.6 mm. Compared to previously proposed spiral coil structures, this expanded planar design facilitates easier tracing of the surfaces. When encountering surface irregularities, the pins within the structure move continuously. The "coil continuum" was fabricated using a 3D printer, and ABS resin was chosen as the material.

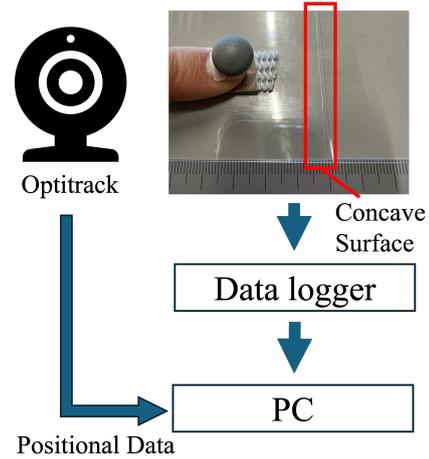

Fig.2  Measurement system

## 3  EXPERIMENT ON DETECTION ACCURACY

### 3.1  Experiment

To evaluate the performance of the developed device, we conducted an experiment to trace concave surfaces with varying widths. Fig. 2 illustrates a schematic of the concave surfaces made from acrylic plates with a thickness of 0.5 mm. We prepared five types of concave surfaces with widths ranging from 1 mm to 5 mm, in 1 mm increments. The concave surfaces were positioned 30 mm from the initial sensor position. A 0.2 mm thick acrylic plate was placed on top to smooth the shape of the concave surfaces. As the device passed over the concave surfaces, the overlaid acrylic plate bent, creating an indentation that the device detected. The strain and position were measured for each condition. To capture the characteristic waveform, five measurements were conducted for each condition and the average value was calculated.

A schematic of the experimental setup is shown in Fig. 3. A data logger (CTES-100A, Kyowa Electronic Instruments) was used to measure strain. Position measurements were performed using motion tracking with an OptiTrack system (V120, Acuity Inc.). A marker was placed on the central part of the device to enable tracking. The sampling frequency was set to 120 Hz for OptiTrack and 1 kHz for the strain gauge. The sampling frequency was set to 120 Hz for OptiTrack and 1 kHz for the strain gauge. At the start of the measurement, the developed device was tapped with an index finger to apply force and ensure a constant sensor pressure value. Once the sensor value stabilized, the device was traced from one end of the evaluation surface to the other with consistent pressure. Subsequently, the data obtained from OptiTrack was linearly interpolated to match a 1 kHz sampling frequency. The timing of the tap with the index finger served as the reference point for synchronizing the data obtained from the sensor.

### 3.2 Experimental Result

Figs. 4–8 present the results obtained for each condition. The vertical axis represents the strain, and the horizontal axis represents the distance from the reference point. The standard deviation (SD) of the strain is plotted as a shaded band in the graph. The waveforms obtained for each condition exhibited characteristic patterns in the range 30–50 mm. Previous research [8] has reported that when tracing an indented surface using a spiral coil structure, a positive peak is followed by a negative peak and another positive peak, with the absolute value of the central peak being the largest. The results obtained in this study exhibited similar characteristics. By analyzing the results for each condition, it is evident that as the concave surface width decreases, the absolute value of the signal peaks also decreases. This is because the depth at which the device penetrated the concave surface became shallower, resulting in less deformation of the structural part.

Next, we discuss the relationship between the signal and device. The 10 mm point indicates the position of the device just before the concave surface. The 26 mm point indicates the position of the strain gauge in the sensor immediately before the concave surface. This is due to the fact that the strain gauge, as shown in Fig. 1, is positioned 4 mm on either side of the center of the device. The points at 39 mm, 38 mm, 37 mm, 36 mm, and 35 mm for the conditions with decreasing concave surface widths indicate that the strain gauge has passed the concave surface. Similarly, the points at 55 mm, 54 mm, 53 mm, 52 mm, and 51 mm for the conditions with decreasing concave surface widths indicate that the device has passed the concave surface.

### 4 Discussion

We evaluated whether the developed device could detect indentations under various conditions based on the obtained results. All results exhibited waveform characteristics consistent with those described in previous research [8], indicating successful detection of indentations across all conditions. However, it is important to note that the "coil continuum" used in the device's structure has a diameter of 1 mm, which may limit its ability to detect indentations smaller than this diameter.

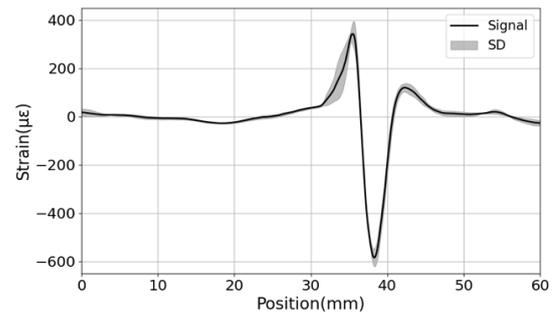

Fig.4 Signals Corresponding to a 5mm Width

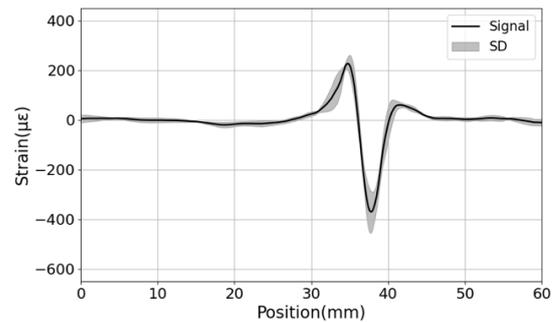

Fig.5 Signals Corresponding to a 4mm Width

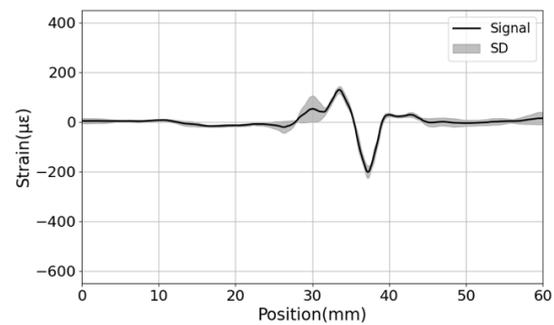

Fig.6 Signals Corresponding to a 3mm Width

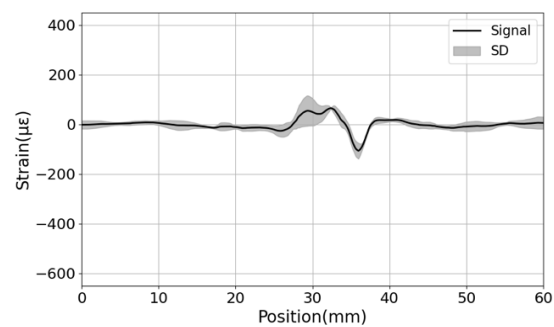

Fig.7 Signals Corresponding to a 2mm Width

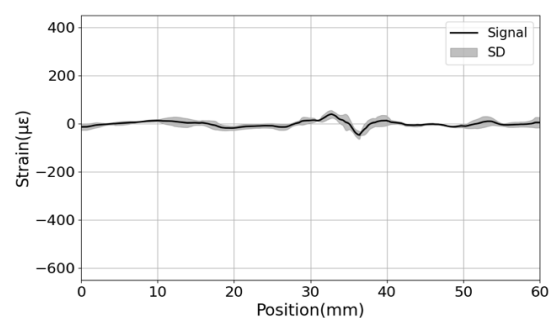

Fig.8 Signals Corresponding to a 1mm width

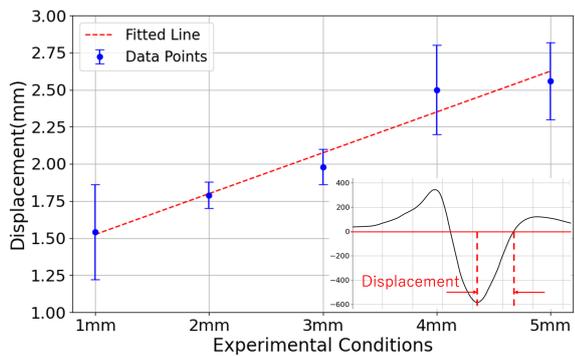

Fig.9  Displacement from peak to rise

Next, we examined whether the device could quantitatively evaluate the indentation widths. To assess the feasibility of quantitative evaluation, we calculated the displacement from the central peak to the point where the strain value reached zero under each condition. When the strain value reaches zero, the strain gauge is no longer deformed, indicating that the device has fully passed over the indentation. Therefore, the displacement from the central peak to the point where the strain value reaches zero may correspond to the width of the indentation.

Fig. 9 presents the results of the displacement calculation from the central peak to the point where the strain value reached zero. The vertical axis represents the displacement, while the horizontal axis represents the conditions of the grooves. Standard deviations are plotted as error bars for each data point. The red dashed line represents the approximation based on the data. As shown in Fig. 9, the displacement increases with the indentation width, suggesting that the signal obtained from the device could be used to quantitatively evaluate the width of indentations.

## 5 Conclusion

In this study, we aimed to automate and systematize surface inspection processes commonly conducted in the automotive industry. To achieve this, we developed a novel tactile device based on the application of a spiral coil structure.

In addition to detecting minute distortions, automating the surface inspection requires automating the action of tracing the surface of an object while applying an appropriate force. Although humans can easily trace the shape of a surface, it is a difficult task for a robot. In this study, we demonstrated the tactile components of a robot for automated surface distortion inspection. In future work, it will be necessary to equip robots with tactile sensors to trace the shapes of their surfaces. To evaluate the performance of the developed device, we conducted experiments by tracing concave surfaces with widths ranging from 1 mm to 5 mm. The experimental results demonstrated that the device could detect concave surfaces under all the conditions.

In the future, the development of a robot capable of tracing the target surface with consistent pressure will be necessary. By equipping the robot with the device proposed in this study, the automation of surface inspection can be realized. Furthermore, detailed analysis of the waveforms obtained from the sensor may enable the quantitative evaluation of various parameters related to micro bumps.

## Acknowledgement

This work was supported by JSPS KAKENHI Grant Numbers 22K14227, 24K00854.